\documentclass[11pt,a4paaper,oneside]{article}
\usepackage{lineno,hyperref}

\usepackage{graphicx}              
\usepackage{epsfig}
\usepackage{amsmath}               
\usepackage{amssymb}
\usepackage{epstopdf}
\usepackage{verbatim}
\usepackage{mathptmx}
\usepackage{mathalpha}
\usepackage{mathtools}
\usepackage[scale=0.80]{geometry}
\usepackage{graphicx, times}
\usepackage{amsfonts}              
\usepackage{amsthm}                
\usepackage{multicol}
\usepackage{algorithm}
\usepackage{algorithmic}
\usepackage{varwidth}
\usepackage{parskip}
\usepackage{hyperref}
\usepackage{rotating}
\usepackage[numbers,sort&compress]{natbib}
\usepackage{multirow}
\usepackage{pdflscape}
\usepackage[numbers]{natbib}
\usepackage{caption}
\usepackage{xcolor}
\usepackage{color}
\usepackage{comment}
\usepackage{subcaption}
\newcommand*{\rom}[1]{\expandafter\@\romannumeral #1}

\newcommand{\bea}{\begin{eqnarray}}
	\newcommand{\eea}{\end{eqnarray}}
\newcommand{\bee}{\begin{eqnarray*}}
	\newcommand{\eee}{\end{eqnarray*}}

\begin{document}
\author{ Romanshu Garg$^{1}$\footnote{romanshugarg18@gmail.com}, Tanmoy Chowdhury $^{2}$\footnote{tanmoych.ju@gmail.com}, G. P. Singh$^{1}$\footnote{gpsingh@mth.vnit.ac.in}, Farook Rahaman $^{2}$\footnote{rahaman@associates.iucaa.in}
\vspace{.3cm}\\
${}^{1}$ Department of Mathematics,\\ Visvesvaraya National Institute of Technology, Nagpur, 440010, Maharashtra India.
\vspace{.3cm}
\\${}^{2}$ Departments of Mathematics, \\ Jadavpur University, Kolkata 700032, West Bengal, India.
\date{}}

\title{Cosmological model with Gong-Zong Parametrization in $f(R,L_m)$ gravity}

\maketitle
\begin{abstract}
We present the cosmic expansion scenarios in the $f(R, L_m)$ gravity studied by using the dark energy equation of state (EoS) parameters. We proceed with the specific form of $f (R, L_m)$ gravity termed as $f (R, L_m)=\frac{R}{2}+L_{m}^{\alpha}$. We derive the expansion rate in terms of the red-shift for two different forms of EoS parameter. In first model, EoS parameter varies inversely with the redshift and in second model, it involves the exponential form with the redshift. By using the Bayesian methods based on the $\chi^{2}$-minimization technique, the median values of model parameters are determined for the cosmic chronometer(CC) and Joint (CC+Pantheon) data sets. The behavior of fundamental cosmological parameters such as the deceleration parameter, energy density and pressure are thoroughly examined. Additionally, the nature of Statefinder diagnostics and the present age of universe exemplifies the compatibility with the late-time astronomical observations.
\end{abstract}

\section{Introduction}\label{sec:1}
Astronomical studies~\cite{1998AJ....116.1009R,1999ApJ...517..565P,2020A&A...641A...6P} have confirmed the expanding nature of the cosmos, which is an intriguing discovery of the observable universe. Afterwards, theoretical cosmologists became interested in creating cosmological models that depicted an accelerated phase of the universe's expansion. Multiple theoretical cosmologists have put forward models that either modify Einstein's field equations or suggest other theories of gravity in response to the universe's acceleration. A more appropriate cosmological models to explain the current growth in the universe's expansion has been proposed based on the dark energy. It is generally agreed that the dynamical cosmological term $\Lambda$ is one of the most promising dark energy candidates~\cite{weinberg1989cosmological}. However, there are multiple difficulties\cite{di2021realm,
weinberg1989cosmological,carroll2001cosmological} with the $\Lambda$ term model, including coincidence issues and fine tunning\cite{carroll2001cosmological}. To deal with these present cosmological problems, various approaches have been proposed in the last few decades. Nowadays, the modified theory of gravity technique is the most often used option for resolving current cosmological issues. One of the most well-known ideas to the dark content problem of the universe is the $f(R)$ gravity~\cite{1982GReGr..14..453K, buchdahl1970non} which is a modification of General Relativity. With the evolving time, several other modified theories are also widely explored\cite{nojiri2011unified, nojiri2017modified,harko2011f, cai2016f, capozziello2011cosmography,
capozziello2019extended,bamba2010finite,
capozziello2023role,singh2020study,singh505abc, garg2025cosmic,goswami2024flrw,
hulke2020variable,singh2024dynamical,
singh2024conservative,
chowdhurystudy,singh2025observational,
kotambkar2017anisotropic,
singh1997new,varela2025cosmological,
singh2022cosmic,singh2022cosmological, singh2022lagrangian, singh2023homogeneous, lalke2023late}. 
\par A relation between the Ricci scalar $R$ with $L_m$ denoting the matter Lagrangian density has given rise to the gravitational theory based on $f(R, L_m)$ and is termed as the $f(R,L_m)$ theory. Harko and Lobo~\cite{harko2010f} show that this theory provides a more general relation between matter and geometry sector. However, this approach violates the equivalence principle and the solar system observations impose stringent restrictions on these models ~\cite{Faraoni2004pi,zhang2007behavior,bertolami2008general,Rahaman2009solar}. This theory belong to the class of the most complete gravitational theories represented on Riemannian space~\cite{nojiri2004gravity,allemandi2005dark} and, offers novel perspectives on the intricate relationships between the geometries of cosmology- astrophysics and matter. Due to the presence of matter, the field equations of $f(R, L_{m})$ gravity and the $f(R)$ model differ from each other and GR, but they coincides in empty space \cite{harko2010f,lobo2015extended}. In connection with the $f(R, L_m)$ gravity, numerous further studies have been completed  \cite{myrzakulov2024linear,bhardwaj2025cosmological,jaybhaye2024late,
myrzakulova2024investigating,garg2024cosmological,
doi:10.1142/S0219887823501050,shukla2023flrw,
maurya2024bianchi,shukla2023dynamical,
jaybhaye2022cosmology}.
\par In the present paper, we probe the dynamics of dynamical dark energy in late universe of $f(R, L_{m})$ gravity with the parametrized EoS parameters. The considered EoS parameter of dark energy varies inversely with the redshift with form $\omega = \frac{w_o}{z+1}$. In another case, it possess exponential form given by $\omega=\frac{w_o}{z+1} \exp{\left( \frac{z}{z+1} \right)}$. These parametrization of dark energy are also termed as the Gong-Zhang parametrization \cite{gong2005probing,
yang2023latest,castillo2023exponential}. These DE EoS parameters shown to be suitable for the description of dark energy characteristic during present era \cite{gong2005probing}. The Gong-Zhang models are distinguishable from the $\Lambda$CDM model based on the growth factor of dark energy \cite{yang2023latest}. The future deceleration scenario will naturally be identified during the universe evolution in exponential form of Gong-Zhang EoS parameter model  \cite{castillo2023exponential}. There is only one parameter in these dark energy EoS parameters and, the dark energy does not diverge in the cosmological history according to these parameters. This well-behaved nature of dark energy in these parametrization makes them appropriate to model the deceleration-acceleration cosmic history. Further, the parameter $w_o$ enables us to trace the present day EoS parameter. The cosmic history described by these parameters further enable us to describe the compatibility of expansion rate of the $f(R,L_m)$ theory. The cosmological features may further be scrutinized based on the observational data.  
We investigate these EoS parameters in the outline of the $f(R,L_m)$ cosmological model.
\par This work has been divided into seven sections: In Sec. (\ref{sec:2}), an overview of the $f(R, L_m)$ gravity theory is briefly discussed. The field equations for the FLRW metric in $f(R, L_{m})$ gravity are presented in Sec. (\ref{sec:3}). In Sec. (\ref{sec:4}), the Hubble parameters in terms of red-shift using dark energy EoS parameters are derived. These parameters enable to provide the observational compatibility in Sec. (\ref{sec:5}). In particular, we derive the median values of model parameters by using the Bayesian analysis technique for Cosmic Chronometer (CC) data and a combined data set composed of CC and supernovae Type Ia observations. The physical and geometrical analysis based on the deceleration parameter and the physical parameters such as (pressure and energy density) are studied in Sec. (\ref{sec:6}). Furthermore, the statefinder diagnostic is used to examine the nature of dark energy based on the geometrical parameters. The universe age in these models are computed to describe the compatibility with Planck based results of $\Lambda$CDM model. The conclusions are given in the last section (\ref{sec:7}). 
\section{An Overview of $f(R, L_{m})$ gravity theory}\label{sec:2}
In this section, we discuss a overview of $ f(R, L_{m})$ gravity and the action in $f(R,L_m)$ gravity takes the form~\cite{harko2010f}.
\begin{equation}{\label{1}}
S=\int f(R, L_{m})\sqrt{-g} d^{4}x,
\end{equation}
where, $ L_{m} $ defines as matter-Lagrangian density and $R$ refers to Ricci scalar. We can define Ricci scalar $(R)$ by using the Ricci tensor ($R_{\mu \nu}$) with metric tensor ($g^{\mu \nu}$) as 
\begin{equation}{\label{2}}
R=g^{\mu \nu}R_{\mu \nu}, 
\end{equation}
where the $R_{\mu \nu}$ is 
\begin{equation}{\label{3}}
R_{\mu \nu}= \partial_{c} \Gamma^{c}_{\mu \nu}-\partial_{\mu} \Gamma^{c}_{c\nu}+\Gamma^{c}_{\mu \nu}\Gamma^{d}_{dc}-\Gamma^{c}_{\nu d} \Gamma^{d}_{\mu c},
\end{equation}
and $ \Gamma^{\alpha}_{\beta \gamma} $ signifies the Levi-Civita connection given by
\begin{equation}{\label{4}}
\Gamma^{\alpha}_{\beta \gamma}= \frac{1}{2}g^{\alpha c}\left(\frac{\partial g_{\gamma c}}{\partial x^{\beta}}+\frac{\partial g_{c \beta}}{\partial x^{\gamma}}-\frac{\partial g_{\beta \gamma }}{\partial x^{c}} \right).
\end{equation}
Using the action (\ref{1}) with the metric tensor $ g_{\mu\nu} $, the field equation may be written as,
\begin{equation}{\label{5}}
\frac{\partial f}{\partial R}R_{\mu \nu}+(g_{\mu \nu} \square -\nabla_{\mu}\nabla_{\nu})\frac{\partial f }{\partial R}-\frac{1}{2}\left( f-\frac{\partial f}{\partial L_{m}}L_{m}\right)g_{\mu \nu}=\frac{1}{2}\left(\frac{\partial f}{\partial L_{m}}\right)T_{\mu \nu}, 
\end{equation}
where $T_{\mu \nu}$ is defined as the energy-momentum tensor for perfect fluid as
\begin{equation}{\label{6}}
T_{\mu \nu}=\frac{-2}{\sqrt{-g}}\frac{\delta(\sqrt{-g}L_{m})}{\delta g^{\mu \nu}}.
\end{equation}
Using these field equations, we can construct a relationship between the Ricci scalar$(R)$, the trace of the energy-momentum tensor  $T$ and the matter Lagrangian density $(L_{m})$ as
\begin{equation}{\label{7}}
R\left(\frac{\partial f}{\partial R}\right)  +2\left(\frac{\partial f}{\partial L_{m}}L_{m}-f\right)+ 3\square \frac{\partial f}{\partial R}=\frac{1}{2}\left(\frac{\partial f}{\partial L_{m}}\right)T, 
\end{equation}
where the $ \square I=\frac{1}{\sqrt{-g}}\partial_{\mu}(\sqrt{-g}g^{\mu \nu} \partial_{\nu}I)$ for any type of random function I. To analyse the Eq. (\ref{5}), we can substitute the covariant derivative with energy-momentum tensor, written as:
 \begin{equation}{\label{8}}
\nabla^{\mu}T_{\mu \nu}=2\nabla^{\mu} \log[f_{L_{m}}(R, L_{m})] \frac{\partial L_{m}}{\partial g^{\mu \nu}}.
\end{equation}
In the next section, we write the equations of motions in the FLRW spacetime.  
\section{ Motion equations in $f(R, L_{m})$ \text{ gravity } }\label{sec:3}
The observable universe is homogeneous and isotropic to a great extent. In order to describe this nature of universe, we employ the flat Friedman–Lemaître–Robertson–Walker (FLRW) metric~\cite{partridge2004introduction} as:
\begin{equation}{\label{9}}  
ds^{2}=a^{2}(t) \left( dx^{2}+ dy^{2}+ dz^{2}\right)-dt^{2},
\end{equation}
where at a given time $t$, the scale factor describing the cosmic evolution is denoted by $a(t)$. For the line element (\ref{9}), the non-vanishing components of the Christoffel symbols are:
\begin{equation}{\label{10}}
\Gamma^{0}_{pq}= -\frac{1}{2}g^{00} \  \frac{\partial g_{pq}}{\partial x^{0}}, \  \  \ \ \Gamma^{r}_{0q}=\Gamma^{r}_{q0}= \frac{1}{2}g^{r\lambda} \  \frac{\partial g_{q \lambda}}{\partial x^{0}},
\end{equation}
where the variables $p$, $q$, $r$ = 1, 2, 3. By utilizing equation (\ref{3}), the non-vanishing components of the Ricci tensor and, the Ricci scalar are written as:
\begin{equation}{\label{11}}
R^{0}_{0}=3\frac{\ddot{a}}{a}, \  \ R^{1}_{1}=R^{2}_{2}=R^{3}_{3}=\frac{\ddot{a}}{a}+2\left(\frac{\dot{a}}{a}\right)^{2}, \quad R=6 \ \left(\frac{\dot{a}}{a}\right)^{2}+6\left( \frac{\ddot{a}}{a}\right) =12H^{2} +6\dot{H},            
\end{equation}
where $ H=\frac{\dot{a}}{a} $ denotes the Hubble parameter describing the expansion rate. For the perfect fluid, we take the  energy-momentum tensor as
\begin{equation}{\label{13}}
T_{\mu \nu} = (\mathit{p} + \rho) u_{\mu} u_{\nu} + p g_{\mu\nu},
\end{equation}
where $\mathit{p}$ and $\rho$ are defined as the isotropic pressure and the energy density of the cosmic fluid and the components of the four velocity are $u^{\mu} = (1, 0, 0, 0)$ satisfying $u_{\mu} u^{\mu} = -1 $. The Friedmann equations that describe the dynamics relation of the universe in  $ f(R, L_m)$ gravity can be read as follows:
\begin{equation}
	\frac{1}{2} (f -f_{L_m} L_m - f_R R) + 3H  \dot{f}_R +3 H^2 f_R  = \frac{1}{2} f_{L_m} \rho,\label{14}
\end{equation}
\begin{equation}
	3H^2 f_R +\dot{H} f_R- \ddot{f}_R - 3H \dot{f}_R + \frac{1}{2} (- f+f_{L_m} L_m ) = \frac{1}{2} f_{L_m} p.\label{15}
\end{equation}
These field equations are useful to derive the expansion rate expression for the description of cosmological history and the transiting universe evolution. 
\section{The expansion rate and background dynamics in the $ f(R, L_{m})$ model}
\label{sec:4}
In this section, we choose a specific form of $ f(R, L_{m})$ gravity~\cite{harko2014generalized} described by:
\begin{equation}{\label{16}}
f(R, L_{m})=\frac{R}{2}+ L_{m}^{\alpha},
\end{equation}
where $ \alpha $ is an arbitrary constant. For this specified functional form of the $f(R, L_{m})$ model with $L_{m}=\rho$~\cite{harko2015gravitational},\\
a combination of equations (\ref{14}) and (\ref{16}) yield
\begin{equation}{\label{17}}
3  H^2= (2 \alpha -1) \rho ^\alpha,
\end{equation}
further, using  $R=12H^{2}+6\dot{H}$, equations (\ref{15}) and (\ref{16}) give
\begin{equation}{\label{18}}
2\dot{H} +3  H^2= (\alpha-1) \rho ^\alpha- \alpha p \rho ^{\alpha -1}
\end{equation}
Mathematically, the Equation of State (EoS) parameter is given by $\omega=\frac{p}{\rho} $. We use equations (\ref{17}), (\ref{18}) and $ \dot{H}=-H(1+z)\frac{dH}{dz}$ to write the EoS parameter $ \omega $ as 
\begin{equation}{\label{20}}
\omega=\frac{2 (2 \alpha -1) (z+1) H'-3 \alpha H}{3 \alpha H}.
\end{equation} 
Here, we have two independent equations (\ref{17}) and (\ref{18}) with three unknown parameters, $a(t)$, $\rho$ and $p$. To derive a specified solution, we may need an additional equation to solve equation (\ref{20}). We utilize the reciprocal- and exponential-like EoS parameters~\cite{gong2005probing} to solve equation (\ref{20}).  Thereafter, now the number of unknowns and the number of equations are same.
\vspace{0.2cm} \\
In the present study, we consider the Gong-Zong parametrizations, (reciprocal like EoS parameter $\omega (z)=\frac{w_{0}}{(1+z)}$ and exponential like EoS parameter $\omega (z)=\frac{w_{0}}{(1+z)}\exp\left( \frac{z}{z+1}\right))$ due to following reasons which may address specific limitations of other dark energy (DE) equation of state (EoS) parametrizations:
\begin{itemize}
\item In contrast to other parametrizations (such as the linear $\omega(z)=w_{0} +w_{1}z$\cite{huterer2001probing}, logarithmic $\omega(z)=w_{0} +w_{1}ln(1+z)$\cite{efstathiou1999constraining}, polynomial $\omega(z)=-1+\alpha(1+z)+\beta(1+z)^{2}$\cite{weller2002future}), the Gong-Zong forms do not diverge at high red-shifts $(z>>1)$.
\item The Gong-Zong parametrization $\omega (z)=\frac{w_{0}}{(1+z)}\exp\left( \frac{z}{z+1}\right)$ approaches to $0$ when $z \to -1$, while other EoS parameters like CPL $\omega = w_{0}+ w_{1}\left( \frac{z}{1+z}\right)$\cite{linder2003exploring}, Generalized JBP $\omega = w_{0}+ w_{1}\left( \frac{z}{(1+z)^{n}}\right)$\cite{liu2008revisiting} are diverging in the asymptotic limit $z \to -1$.
\item Both of these Gong-Zong EoS parameter forms are depending on a single parameter $w_{0}$, which may directly represents the present-day DE EoS value $\omega(z=0)=w_{0}$. This simplicity helps to reduces degeneracy in the observational constraining. The other forms of EoS parameters (like the CPL, polynomial, logarithmic, Generalized JBP) have more than one model parameter.
\end{itemize}
The considered forms of EoS parameters are chosen for their mathematical regularity, minimal parameters and ability to describe the observed transition from deceleration to acceleration without theoretical pathologies. The considered Gong-Zong parametrizations having one parameter may enable us to study the geometry of universe in simplified way with other density parameters. In these models, the dark energy may track the matter in past. These models may even have the past of matter-dominated evolution. These characteristics may induce quite important implications in the late universe evolution. Due to these important and interesting properties, the Gong-Zhang form of EoS parameters may provide interesting implications of the universe evolution in the $f(R,L_m)$ gravity.

\subsection{Model-I}
In this model, we investigate with reciprocal-like equation of state parameter~\cite{gong2005probing},
\begin{equation}{\label{101}}
\omega  =\frac{w_o}{z+1},
\end{equation}
where, we may use the observations to constrain the parameter $w_0$. In the outline of cosmological modelling, this EoS parameter
offers several important advantages. First of all, it offers a framework that can be physically understood with parameter $w_0$. This parameter is denoting the DE EoS parameter at the current epoch $(z=0)$. The inclusion of red-shift dependency enables to characterize the evolution of DE. As $z$ goes more rapidly towards infinity (in the past), \( \omega \sim 0 \), which demonstrates that the DE EoS parameter leans towards zero at early era. As \( z \to -1 \) (in the asymptotic limit), \( \omega \to -\infty \), it indicates that the DE EoS parameter leans towards negative infinity in the far future. These properties of the DE EoS parameter are reasonable and attractive alternative to study its evolution in $f(R, L_m)$ gravity. From equation (\ref{20}) and (\ref{101}), we get
\begin{equation}{\label{96}}
\frac{w_o}{z+1}=\frac{2 (2\alpha -1) (z+1) H'-3\alpha H}{3 \alpha H}.
\end{equation}
By solving equation (\ref{96}), we obtain the Hubble parameter's expression as follows
\begin{equation}{\label{21}}
H(z)= H_0 \exp \left(\frac{3\alpha \left(z w_o+(z+1)\log (z+1)\right)}{2 (2\alpha-1) (z+1)}\right).
\end{equation}
where $H_{0}$ denotes the current value of the Hubble parameter. In the $f(R,L_m)$ gravity, the reciprocal-like EoS parameter yields the Hubble parameter with exponential function and leads to a non-trivial evolution as compared to the $\Lambda$CDM model. It would be interesting to investigate whether this model traces the matter-dominated era of the universe evolution or not.   

\subsection{Model-II}
In this model, we investigate with exponential-like equation of state parameter~\cite{gong2005probing} as
\begin{equation}{\label{102}}
	\omega=\frac{w_o}{z+1} \exp{\left(  \frac{z}{z+1} \right) }
\end{equation}
When $z \gg 1$, $\omega$ $\sim 0$. 
A significant distinction between model (\ref{101}) and model (\ref{102}) is that in model (\ref{102}), $\omega$ approaches $0$ when $z \rightarrow -1$. In the past, these two forms exhibited nearly identical behavior but their future evolution diverges significantly. From equations (\ref{20}), (\ref{102}) and after solving them, we obtain the Hubble parameter with respect to redshift as,
\begin{equation}{\label{103}}
	H(z) = H_0 \exp\left[ \frac{3\alpha \left( \log(1+z) + \left( -1 + \exp \left(\frac{z}{1+z}\right) \right) w_0 \right)}{4\alpha - 2} \right]
\end{equation}
where $H_0$ defined as Hubble parameter value at $z = 0$.

\section{Observational constraints}\label{sec:5}
In this section, we test the observational compatibility of the expansion rates (\ref{21}) and (\ref{103}) by conducting a Bayesian analysis with the observations of cosmic chronometer and supernovae type Ia (Pantheon) sample. We employ the joint data sample and term it as the CC+Pantheon sample. The Bayesian analysis is employed to conduct the $\chi^{2}$ minimization with the Markov Chain Monte Carlo (MCMC) technique. We use the emcee Python library ~\cite{foreman2013emcee} in the present analysis.
\subsection{The observation of Cosmic chronometer dataset}\label{sec:5.1}
\begin{center}
	\begin{figure}
		\includegraphics[width=16.5cm, height=9.5cm]{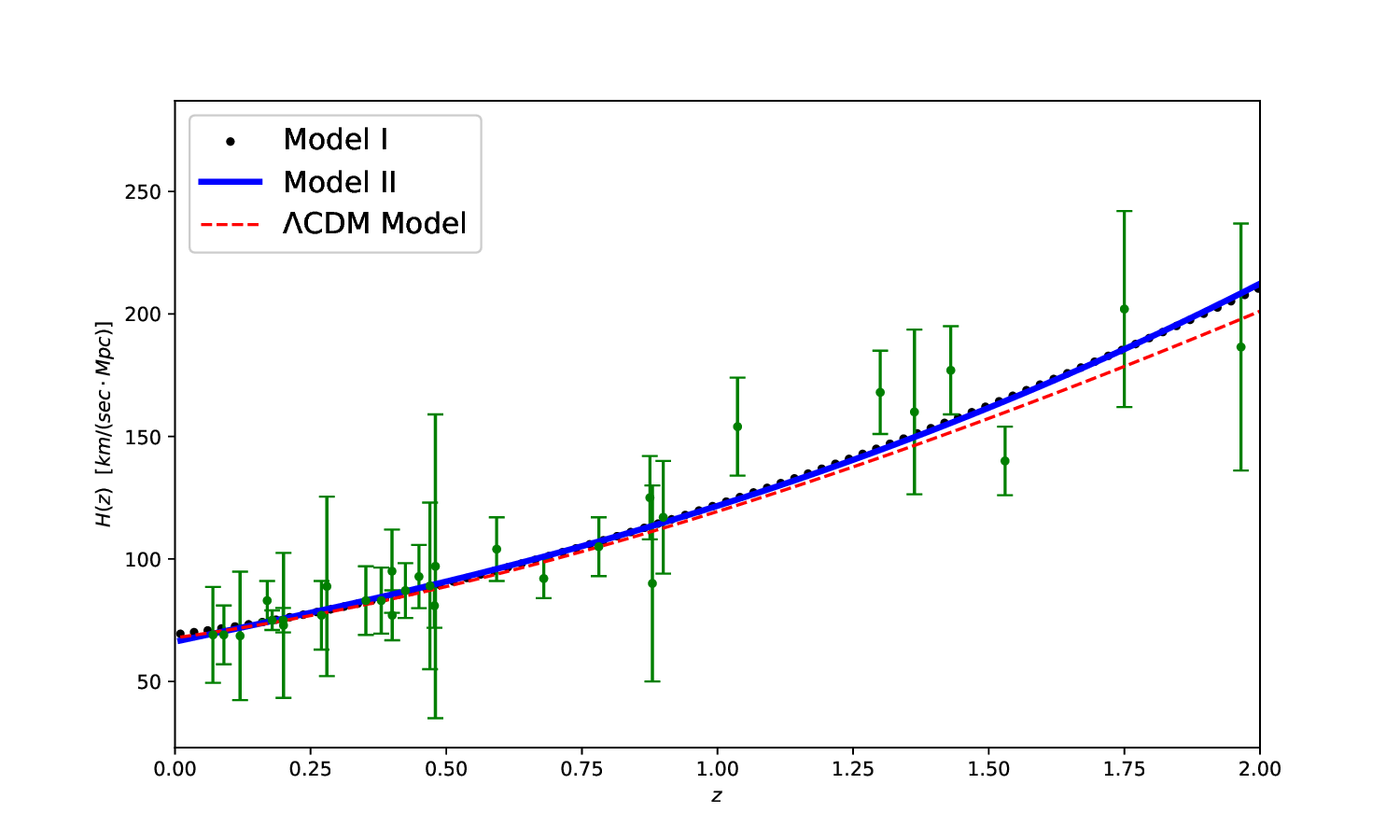}
		\caption{In comparison to the $\Lambda CDM$ model, the best fit Hubble parameters  (given by Eq. (\ref{21}) and (\ref{103})) with $\mathit{z} $.}
		\label{fig:1B}
	\end{figure}
\end{center}

\begin{center}
\begin{figure}
\includegraphics[width=18.5cm, height=19.5cm]{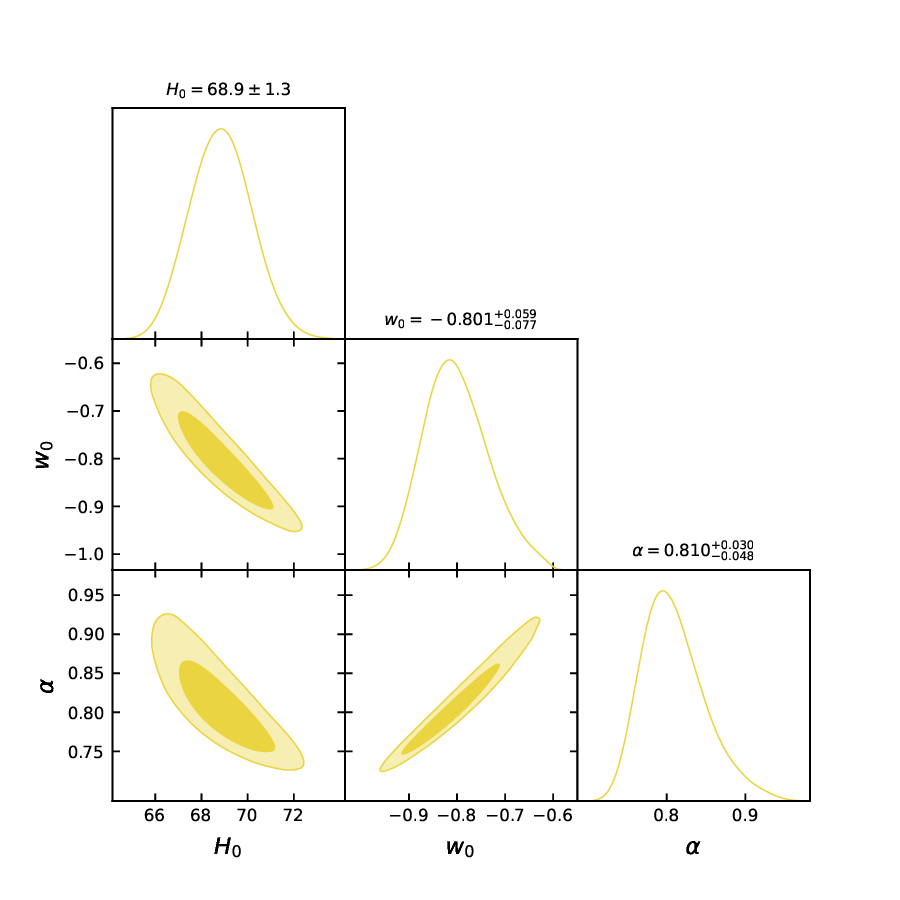}
\caption{Model-I: Marginalized $1\sigma$ and $2\sigma$ contours map with median values of $H_{0}$, $w_{0}$ and $\alpha $ for CC data set.}
\label{fig:2}
\end{figure}
\end{center}

\begin{center}
	\begin{figure}
		\includegraphics[width=18.5cm, height=19.5cm]{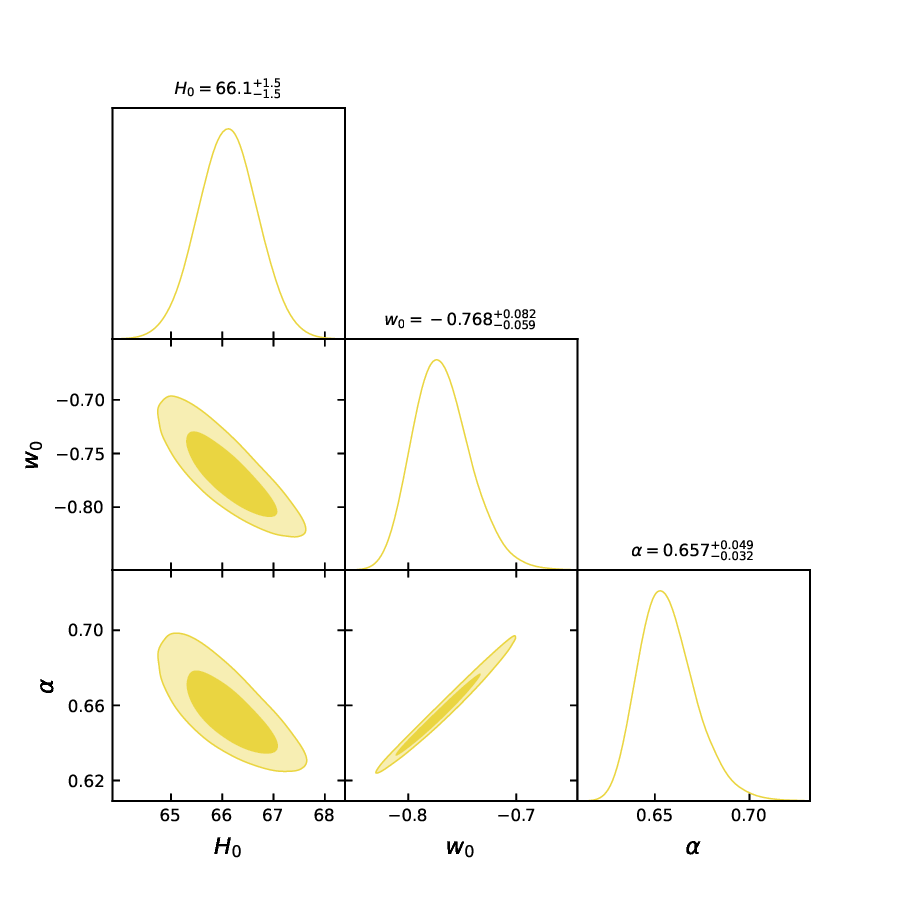}
		\caption{Model-II: Marginalized $1\sigma$ and $2\sigma$ contours map with median values of $H_{0}$, $w_{0}$ and $\alpha $ for CC data set.}
		\label{fig:2B}
	\end{figure}
\end{center}

We analyse the CC data set with 31 observations. The observations are acquired via the differential ages (DA) method of passively evolving galaxies and the range of redshift is $0.07 \leq z \leq 1.965$. This investigation focuses on determining the median values of the model parameters. According to the basic principle set up by Jimenez and Loeb\cite{jimenez2002constraining}, the relationship between the Hubble parameter $(H(z))$, cosmic-time $(t)$, and redshift $(z)$ may be stated as follows: $H(z)=\frac{-1}{(1+z)}\frac{dz}{dt}$. We determine the model parameters $H_{0}, w_{0}$ and $\alpha $ through the minimization of the $\chi^{2}$ function (which is comparable to the maximizing likelihood function) expressed as \cite{mandal2024late, mandal2023cosmic,singh2024affine}.
\begin{equation}{\label{27}}
\chi^{2}_{CC}(\theta)=\sum_{i=1}^{i=31} \frac{[H_{th}(\theta,z_{i})-H_{obs}(z_{i})]^{2}    }{ \sigma^{2}_{H(z_{i})}}.   
\end{equation} 
Here, $H_{th}$ characterizes the theoretical value of Hubble parameter, $H_{obs}$ characterizes its observed value and  $\sigma_{H}$ characterizes the standard error of the observed value. The error between the CC data points and the best-fit Hubble parameter curve is displayed in figure $(\ref{fig:1B})$.
Furthermore, figure $(\ref{fig:2})$ and $(\ref{fig:2B})$ represent a contour map demonstrating the $1\sigma$ and $2\sigma$  likelihood confidence levels for the median values of $H_{0}$, $w_{0}$ and $\alpha $ obtained from the CC data.
\subsection{The observation of  Pantheon dataset}\label{sec:5.2}
\begin{center}
\begin{figure}
\includegraphics[width=18.5cm, height=18.5cm]{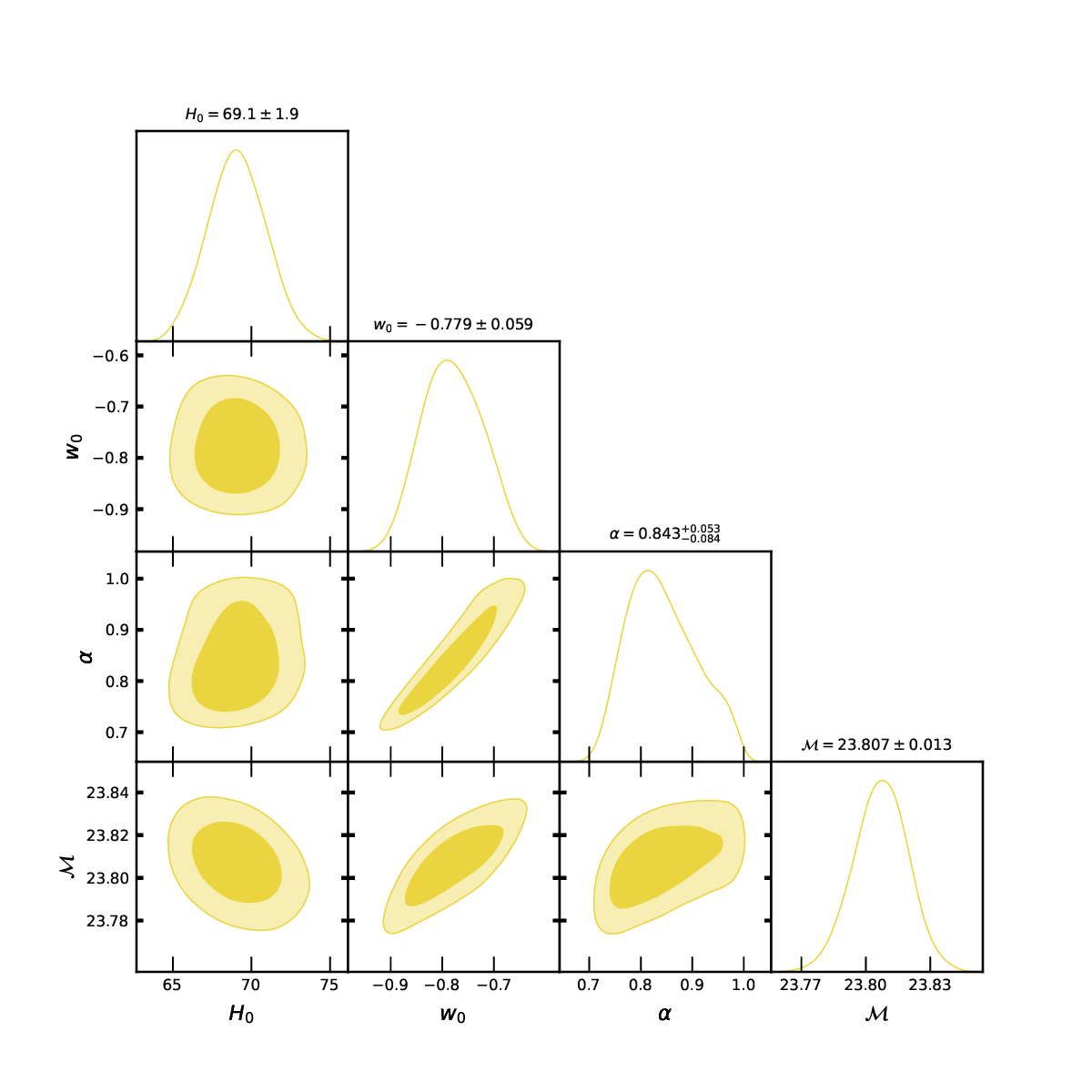}
\caption{Model-I: $1\sigma$ and $2\sigma$  likelihood contours map for the Joint dataset.}
\label{fig:3}
\end{figure}
\end{center}

\begin{center}
	\begin{figure}
		\includegraphics[width=18.5cm, height=18.5cm]{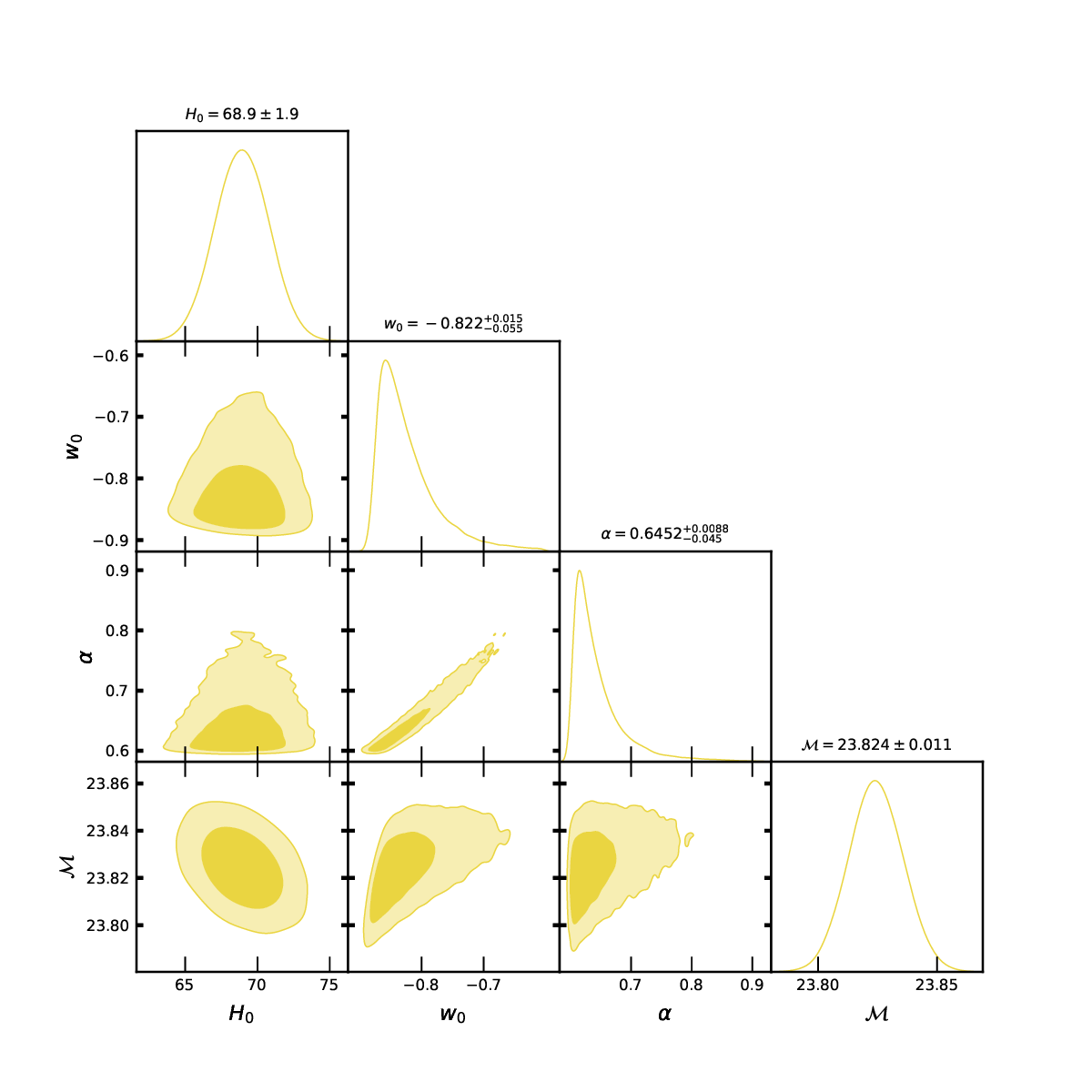}
		\caption{Model-II: $1\sigma$ and $2\sigma$  likelihood contours map for the  Joint dataset.}
		\label{fig:3B}
	\end{figure}
\end{center}

We also utilize the Pantheon dataset, which comprising $1048$ Type Ia supernovae data points for the redshift range $0.01 < z < 2.26$~\cite{scolnic2018complete}. The SNIa sample is composed of the data from surveys such as CfA1-CfA4~\cite{riess1999bvri,hicken2009improved}, Pan-STARRS1 Medium Deep Survey~\cite{scolnic2018complete}, SDSS~\cite{sako2018data}, SNLS~\cite{guy2010supernova}, and the Carnegie Supernova Project (CSP)~\cite{contreras2010carnegie}. \\
The theoretically expected apparent magnitude $\mu_{th}(z)$ is determined by
\begin{equation}{\label{7a}}
\mu_{th}(z)=M+5log_{10}\left[\frac{d_{L}(z)}{Mpc}\right]+25,
\end{equation}
where $M$ is the absolute magnitude. Also, the luminosity distance $d_{L}(z)$ (having dimension of the Length) may be written as~\cite{odintsov2018cosmological}
\begin{equation}{\label{8a}}
d_{L}(z)=c(1+z)\int_{0}^{z}\frac{dz'}{H(z')},
\end{equation}
where the parameter $z$ signifies SNIa's redshift as determined in the cosmic microwave background (CMB) rest frame and $c$ is defined as speed of the light. The luminosity distance $(d_L)$ is often replaced by the dimensionless Hubble-free luminosity distance $(D_{L}(z) \equiv H_{0}d_{L}(z)/c)$. Equation (\ref{7a}) can be written in the following way
\begin{equation}{\label{9a}}
\mu_{th}(z)=M+5\log_{10}\left[D_{L}(z)\right]+5\log_{10}\left[\frac{c/H_{0}}{Mpc}\right]+25. 
\end{equation}
There will be a degeneracy between $H_{0}$ and $M$ in the $\Lambda$CDM model framework~\cite{ellis2012relativistic,asvesta2022observational}.
We  express $\mathcal{M}$ as a combination of these parameters as shown below.
\begin{equation}{\label{10a}}
\mathcal{M}\equiv M+5\log_{10} \left[\frac{c/H_{0}}{Mpc}\right]+25=M-5\log_{10}(h)+42.38, 
\end{equation}
where $H_{0}=h \times 100$ $\frac{Km}{sec.(Mpc)}$. In MCMC analysis, we employ these parameter with appropriate $\chi^{2}$ for the Pantheon dataset as~\cite{singh2024affine,mandal2024late,lalke2024cosmic,
asvesta2022observational,singh2025observational}
\begin{equation}{\label{11a}}
\chi^{2}_{P}= \nabla \mu_{i}C^{-1}_{ij}\nabla \mu_{j},
\end{equation}
whereas $\nabla \mu_{i}=\mu_{obs}(z_{i})-\mu_{th}(z_{i})$, $C_{ij}^{-1}$ defined as the inverse of covariance matrix and $\mu_{th}$ may be specified by equation (\ref{9a}). \\
The luminosity distance is contingent on the Hubble parameter. Following this approach, we utilize the emcee package~\cite{foreman2013emcee} and equations (\ref{21}), (\ref{103}) to obtain the maximum likelihood estimate (MLE) by using the combined of CC+Pantheon data set. We define the joint $ \chi^{2}$ expression as $\chi^{2}_{total}$ = $\chi^{2}_{CC}+\chi^{2}_{P}$ for computing the maximum likelihood estimate using the joint data. Figure $(\ref{fig:3})$ and $(\ref{fig:3B})$ is exhibited for $1\sigma$ and $2\sigma$ likelihood contour maps and 1D distributions with MCMC analysis through combined of CC+Pantheon datasets.
\vspace{0.2cm}\\
The uncertainties (i.e., $1 \sigma$ and $2 \sigma$ confidence intervals) represent the statistical confidence with which the model parameters are determined. Small uncertainties suggest that the model is tightly constrained by the observational data and the predictions of the model (e.g., Hubble rate, deceleration parameter, transition red-shift) are more reliable and precise. Larger uncertainties suggest that degeneracy between parameters or limited constraining power from the data and model predictions may vary significantly within the allowed parameter space, reducing predictive precision.
\vspace{0.2cm}\\
Model I shows relatively tighter confidence intervals for $H_{0}$, $\omega_{0}$ and $\alpha$, especially when constrained by the joint CC + Pantheon dataset. This reflects its better compatibility with observations and greater predictive power. In contrast, Model II yields slightly broader uncertainties, particularly in $\omega_{0}$ and $\alpha$, indicating a weaker constraint from data.
\vspace{0.2cm}\\
For Model I (\ref{21}) and Model II (\ref{103}), the median values of the model parameters computed through MCMC analysis with emcee are provided in tables (\ref{table:1}) and (\ref{table:2}).
\vspace{0.2cm}\\
\begin{table}[htbp]
\centering
\begin{tabular}{|c|c|c|c|c|c|c|c|c|}
\hline
Dataset & $H_{0}$[Km/(sec. Mpc)] & $w_{0}$ & $\alpha$ & $\mathcal{M}$ & $q_{0}$ & $z_t$ & $t_{0}$[Gyr] \\
\hline
CC & $68.9^{+1.3}_{-1.3}$ & $-0.801^{+0.059}_{-0.077}$ & $0.810^{+0.030}_{-0.048}$ &- & $-0.60$ & $0.635$ &$ 13.22^{+0.165}_{-0.34}$  \\
\hline
CC+Pantheon  & $69.1^{+1.9}_{-1.9}$  & $-0.779^{+0.059}_{-0.059}$ &  $0.843^{+0.053}_{-0.084}$ & $23.807^{+0.013}_{-0.013}$ & $-0.59$ & $0.71$ & $13.41^{+0.14}_{-0.46}$ \\
\hline
\end{tabular}
\caption{ {\bf{Model-I:}} Median values of model parameters with the current values of $q_0$, $z_t$ and $t_0$ for CC and joint data sets.}
\label{table:1}
\end{table}
\begin{table}[htbp]
	\centering
	\begin{tabular}{|c|c|c|c|c|c|c|c|c|}
		\hline
		Dataset & $H_{0}$[Km/(sec. Mpc)] & $w_{0}$ & $\alpha$ & $\mathcal{M}$ & $q_{0}$ & $z_t$ & $t_{0}$[Gyr] \\
		\hline
		CC & $66.1^{+1.5}_{-1.5}$ & $-0.768^{+0.082}_{-0.059}$ & $0.657^{+0.049}_{-0.032}$ & - & $-0.2718$ & $0.7$ & $12.74^{+0.18}_{-0.14}$ \\
		\hline
		CC+Pantheon  & $68.9^{+1.9}_{-1.9}$  & $-0.822^{+0.015}_{-0.055}$ &  $0.6452^{+0.0088}_{-0.045}$ & $23.824^{+0.011}_{-0.011}$& $-0.4067$ & $0.871$ & $12.62^{+0.33}_{-0.22}$ \\
		\hline
	\end{tabular}
	\caption{{\bf{Model-II:}} Median values of model parameters with the current values of $q_0$, $z_t$ and $t_0$ for CC and joint data sets. }
	\label{table:2}
\end{table}
\vspace{0.3cm}\\
For Model I (\ref{21}), the median values of the model parameters computed through MCMC analysis with emcee which is mentioned in Table (\ref{table:1}) are compatible with the observational constraint values estimated in \cite{pradhan2023reconstruction,pradhan2023f,doi:10.1142/S0217751X2550099X}. Similarly, for Model II (\ref{103}) the median values of the model parameters which is mentioned in Table (\ref{table:2}) are compatible with the observational constraint values estimated in \cite{doi:10.1142/S0217751X2550099X, yang2018interacting}.

\section{The physical and dynamical characteristics}\label{sec:6}

\subsection{Analysis of Deceleration parameter }\label{sec:6.1}
\begin{figure}[!htb]
\captionsetup{skip=0.4\baselineskip,size=footnotesize}
   \begin{minipage}{0.40\textwidth}
     \centering
     \includegraphics[width=9.0 cm,height=7.5cm]{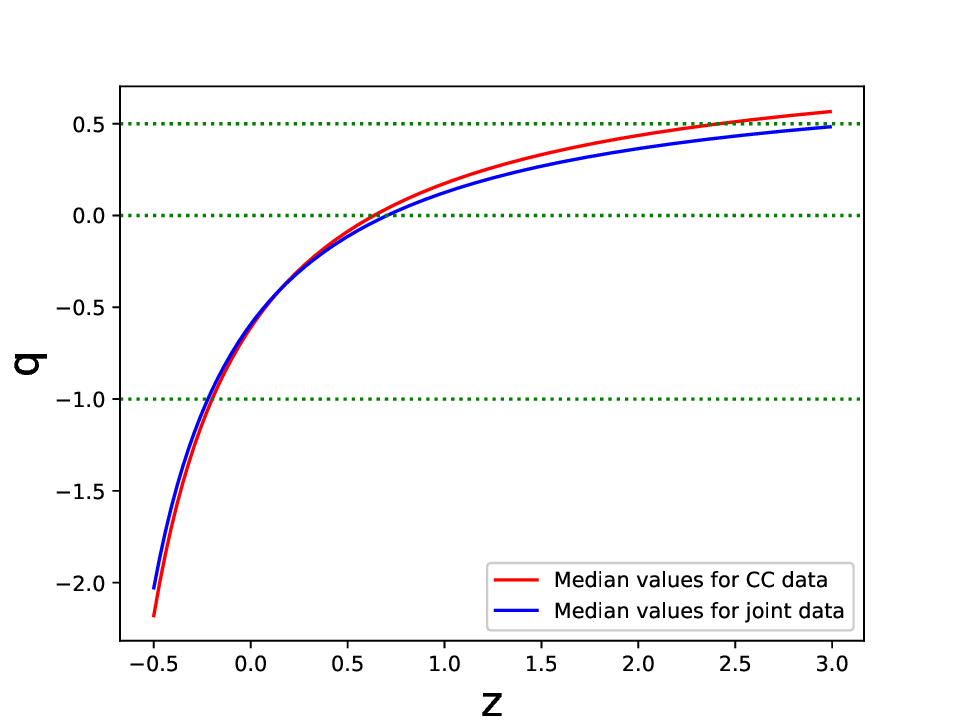}
\caption{ {\bf{For Model-I:}} $q$ with $z$.}
\label{fig:4}
    \end{minipage}\hfill
   \begin{minipage}{0.40\textwidth}
     \centering
     \includegraphics[width=9.0 cm,height=7.5cm]{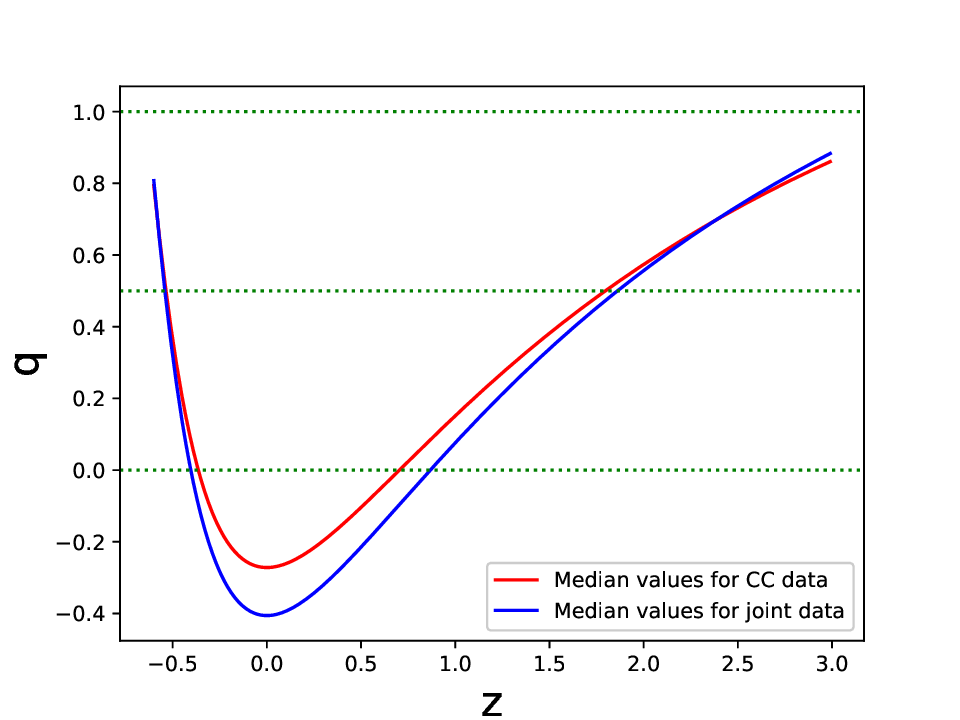}
  \caption{ {\bf{For Model-II}}: $q$ with $z$.}
\label{fig:5}
   \end{minipage}
\end{figure}

The deceleration parameter $(q)$ plays a crucial role in describing the expansion behaviour of the universe. Variations values of the deceleration parameter $(q)$ are used to illustrate the different phases of universe's expansion. The universe expands with acceleration for $q < 0$ and it decelerate for $q > 0$. When the deceleration parameter is smaller than $-1$, the cosmos enters a period of super accelerated expansion. The de Sitter, matter-dominated and radiation-dominated phases of the universe are for $-1$, $\frac{1}{2}$ and $1$ respectively. The deceleration parameter may be expressed as 
\begin{equation}{\label{22}}
 q = -1 + \frac{d}{dt}\frac{1}{H}.
\end{equation}
By using (\ref{21}), (\ref{103}) and (\ref{22}), we obtain for Model I and II as,
\begin{eqnarray}
q(z)=\frac{3\alpha w_{o}-(\alpha-2) (z+1)}{2 (2\alpha-1) (z+1)}, \label{23} \\
q(z) = \frac{-(\alpha -2 )(1+z) + 3\alpha w_0 \exp\left( \frac{z}{1+z}\right) }{2(1+z)(2\alpha - 1)}, \label{104}
\end{eqnarray}
respectively. The analysis of deceleration parameters for Model I and Model II are presented in figures $(\ref{fig:4})$ and $(\ref{fig:5})$. For model I, figure $(\ref{fig:4})$ demonstrates the transition of the universe's evolution from a decelerated expansion phase to an accelerated expansion phase. This transition occurs at $z=0.635$ and $z=0.71$ with median values for CC data and joint data set respectively. For Model I the current  values of deceleration parameter are $q_{0}=-0.6$ (for CC data) and $q_{0}=-0.59$ (for joint data). The negative value signifies that the universe is currently undergoing accelerated expansion. It is evident from the analysis that the deceleration parameter is consistent with present accelerated expansion phase.\\
According to figure $(\ref{fig:5})$, in Model II, the present value of deceleration parameter is $q_0 = -0.2718$ for the CC estimate and $q_0 = -0.4067$ for the Joint estimate. The universe transitions from a decelerated to an accelerated expansion phase at $z=0.7$ and $z=0.871$ for CC and joint data respectively. For red-shift $z < 0.7$, the accelerated phase of cosmic expansion can be observed in Model II. The model explains the accelerated expansion era in the present time, which is consistent with observations. Decelerated expansion phase is predicted by the model in future, which may be possibly due to the energy exchange between dark matter and dark energy. In this scenario, energy is transferred from dark energy to dark matter. Several studies in the literature which predict this type of behavior in the future\cite{castillo2023exponential,chakraborty2014third,escobal2024can}.\\

\subsection{The energy density, pressure and equation of state parameter}\label{sec:6.2}
\begin{figure}
\includegraphics[width=15.5cm,height=8.5cm]{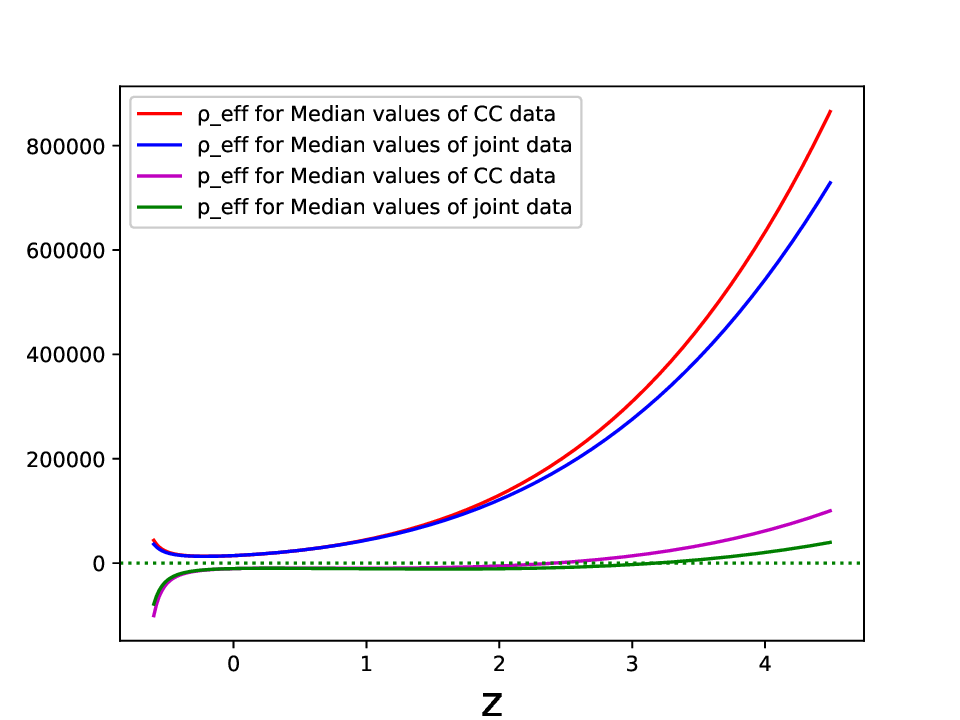}
\caption{ {\bf{For Model-I:}} $\rho_{eff}$ and $p_{eff}$ with $\mathit{z}$  }
\label{fig:6}
\end{figure}

\begin{figure}
\includegraphics[width=15.5cm,height=8.5cm]{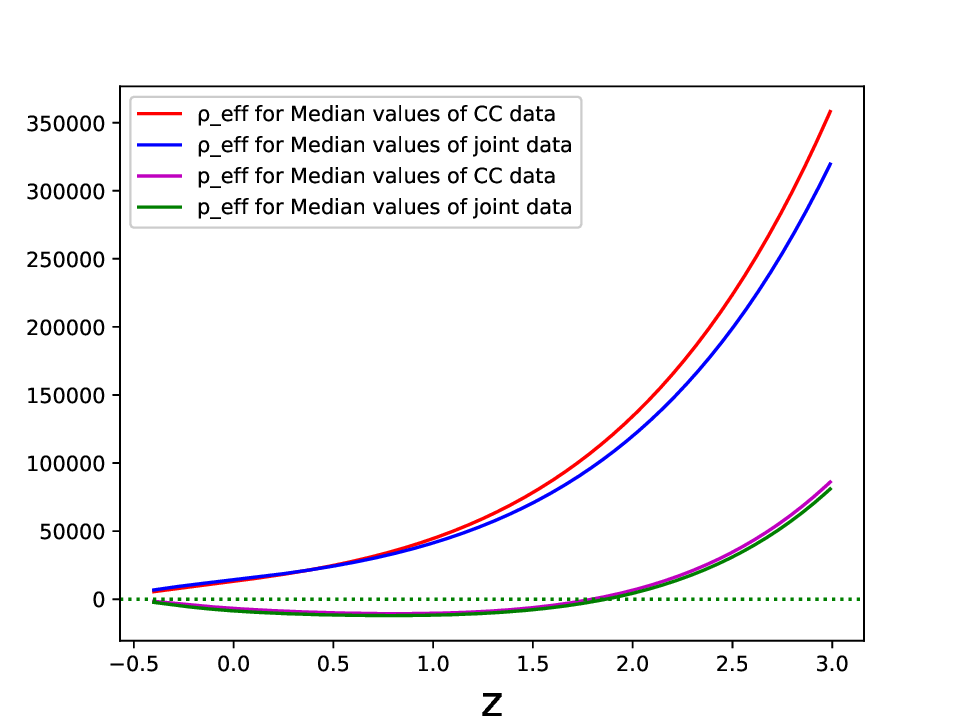}
\caption{ {\bf{ For Model-II:}} $\rho_{eff}$ and $p_{eff}$ with $\mathit{z}$  }
\label{fig:6B}
\end{figure}

\begin{figure}[!htb]
	\captionsetup{skip=0.4\baselineskip,size=footnotesize}
	\begin{minipage}{0.40\textwidth}
		\centering
		\includegraphics[width=9.0 cm,height=7.5cm]{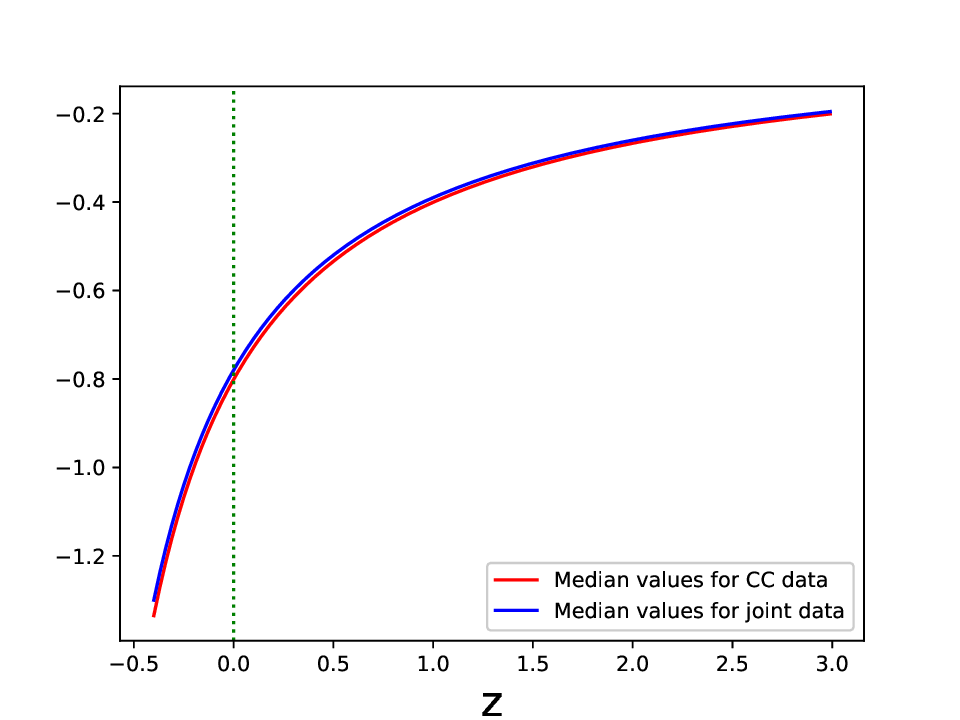}
		\caption{ {\bf{For Model I:}} $\omega$ with $z$.}
		\label{fig:4B}
	\end{minipage}\hfill
	\begin{minipage}{0.40\textwidth}
		\centering
		\includegraphics[width=9.0 cm,height=7.5cm]{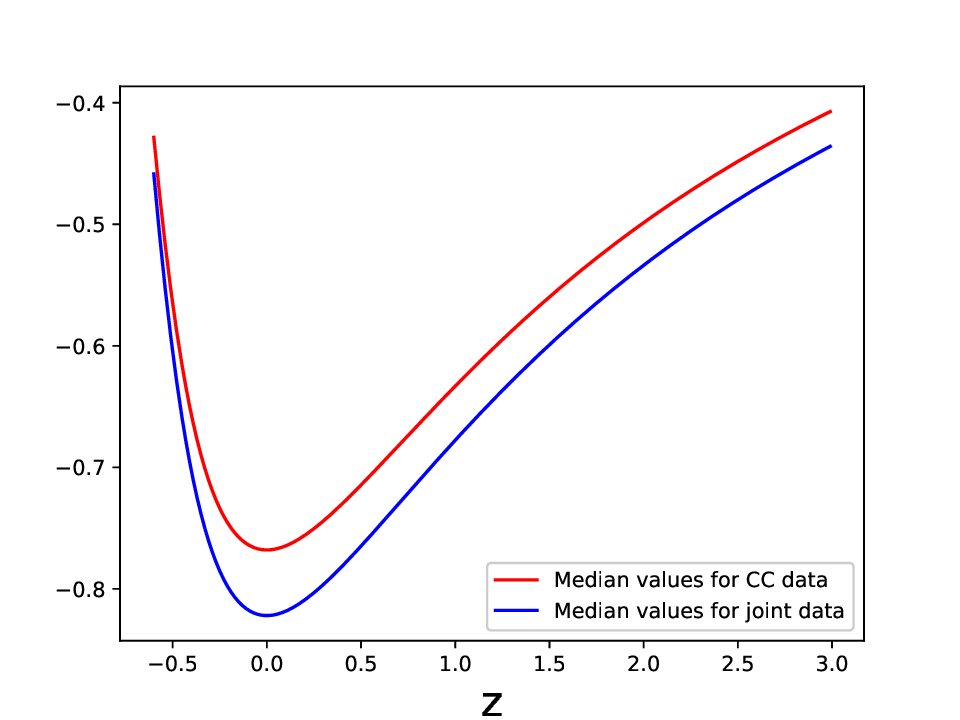}
		\caption{  {\bf{For Model II}} $\omega$ with $z$.}
		\label{fig:5B}
	\end{minipage}
\end{figure}

We study the physical characteristics demonstrated by key quantities such as the energy density and pressure. For the constrained parameters, the energy density is still positive during the cosmic expansion, although the pressure might become negative. For the accelerated expansion of the universe, energy density has been consistently positive while pressure demonstrates negative behavior in the present time and future. Using the equations (\ref{17}), (\ref{21}) and (\ref{103}), energy density in these models may be written as:
\begin{equation}{\label{24}}
\rho(z) = \left( \frac{3}{2\alpha-1}\right) ^{1/\alpha} \left(H_0^2 \exp{ \left(  \frac{3\alpha \left(\log(z+1) + \frac{zw_0}{z+1}\right)}{2\alpha-1} \right) }   \right)^{1/\alpha}	(Model-I)
\end{equation}
\begin{equation}{\label{106}}
	\rho(z) = \left( \frac{3}{2\alpha-1}\right) ^{1/\alpha} \left(H_{0}^{2} \exp\left[ \frac{6\alpha \left( \log(1+z) + \left( -1 + \exp  {  \left(\frac{z}{1+z}\right)   } \right) w_0 \right)}{4\alpha - 2} \right]\right)^{1/\alpha }	(Model-II)
\end{equation}
From using equations (\ref{18}), (\ref{21}), and (\ref{103}), pressure for Model-I and Model-II can be written as:
\begin{equation}{\label{25}}
\mathit{p}(z) = \left[\frac{w_0}{z+1}\right] \left( \frac{3}{2\alpha-1}\right) ^{1/\alpha} \left(H_0^2 \exp  \left(  {\frac{3\alpha \left(\log(z+1) + \frac{z.w_0}{z+1}\right)}{2\alpha-1}}\right) \right)^{1/\alpha}		(Model-I)
\end{equation}
\begin{equation}{\label{107}}
	\mathit{p}(z) = \left[\frac{w_o}{z+1} \exp   \left(\frac{z}{z+1}\right)   \right] \left( \frac{3}{2\alpha-1}\right) ^{1/\alpha} \left(H_{0}^{2} \exp\left[ \frac{6\alpha  \left( \log(1+z) + \left(-1 + \exp\left( \frac{z}{1+z}\right) \right) w_0 \right)}{4\alpha - 2} \right]\right)^{1/\alpha}		(Model-II)
\end{equation} 
For the constrained parameters, the behavior of energy density as well as the pressure are shown in figure (\ref{fig:6}) (for model-I) and figure (\ref{fig:6B}) (for model-II) respectively. These figures demonstrate that the energy densities will increase with red-shift $(z)$ (decrease with cosmic time $(t)$) and remains positive for both the models throughout during expansion. The energy density remains to be positive, supporting the ongoing expansion of the universe. In Model-I, pressure starts with positive values at high red-shift in the early universe while in both current and later epoch, Model-I exhibit negative pressure. These studies are consistent with the accelerating universe’s expanding behavior.
The EoS parameter serves as a vital tool for identifying the nature of dark energy in the cosmological models. The relationship between pressure $(\mathit{p})$ and energy density $(\rho)$ within the universe is described by the EoS parameter. Mathematically, it is defined as $ \omega= \frac{\mathit{p}}{\rho}. $
The dust phase $(\omega=0)$, the radiation-dominated phase $(\omega=\frac{1}{3})$, and the vacuum energy phase $(\omega=-1)$ corresponding with the $\Lambda$CDM model are some of the phases seen through the EoS parameter. Furthermore, there is the accelerated expansion period of the universe for $ (\omega < -\frac{1}{3})$.  
\vspace{0.2cm}\\
Figures $(\ref{fig:4B})$ and $(\ref{fig:5B})$ illustrate the graphical evolution of the EoS parameter for the models. At present $(z=0)$, the estimated values of the EoS parameter are $\omega=-0.801$ for the CC dataset and $\omega=-0.779$ for the joint dataset in Model I. 
In model II, the EoS parameter values are $\omega=-0.768$ and $\omega=-0.822$ at $z = 0$ for CC and joint estimates data respectively.
\vspace{0.2cm}\\
These models possess a quintessence-like dark energy at the present epoch as shown by the behavior of the EoS parameter's figures $(\ref{fig:4B})$ and $(\ref{fig:5B})$. In Model I, $ \omega $ will eventually cross the  $ \omega =-1 $ line, having phantom dark energy dominated behavior in future. The dynamics of Model II indicate a possible energy  exchange between dark matter and dark energy, potentially causing future deceleration era in this model. This feature of the model stands out as particularly intriguing.
\subsection{Energy conditions}\label{sec:6.1}
At a specific point in space-time, the point-wise energy conditions that rely solely on the stress energy tensor are as follows~\cite{singh2022lagrangian,visser1997energy,singh2022cosmological}:
\begin{itemize}
\item[NEC:-] Null Energy Condition $(\rho_{eff} + p_{eff} \geq 0)$ always demonstrates that the addition of energy density and pressure is positive.  
\item[WEC:-] The condition of weak energy indicates that both of the energy density and the sum of pressure-energy density are non-negative.  $(\rho_{eff} \geq 0, \  \ \rho_{eff} + p_{eff} \geq 0) $
\item[DEC:-] Dominant energy condition is denoted by the positivity of energy density ($\rho_{eff}$) and $( \rho_{eff}  \geq \mid p_{eff}\mid ) $.
\item[SEC:-]  Strong energy condition is composed of the conditions $\rho_{eff} + 3p_{eff}\geq 0$ and $\rho_{eff} + p_{eff}\geq 0$.
\end{itemize}
In the decelerating era, the gravitational mass $(\rho_{eff}+3p_{eff})$ continues to be positive. Nevertheless, observational evidence proposes a potential violation of this condition sometimes between the era of galaxy formation and the present time. This variation shows the possible presence of component having negative pressure, and it displays anti-gravitational properties. Meanwhile, the Strong Energy Condition (SEC) contains two inequalities. It is important to remember that the violation of either one will violate the SEC \cite{singh2022lagrangian,singh2023homogeneous,
lalke2023late}.
\par
\begin{figure}
\includegraphics[width=17.5cm,height=8cm]{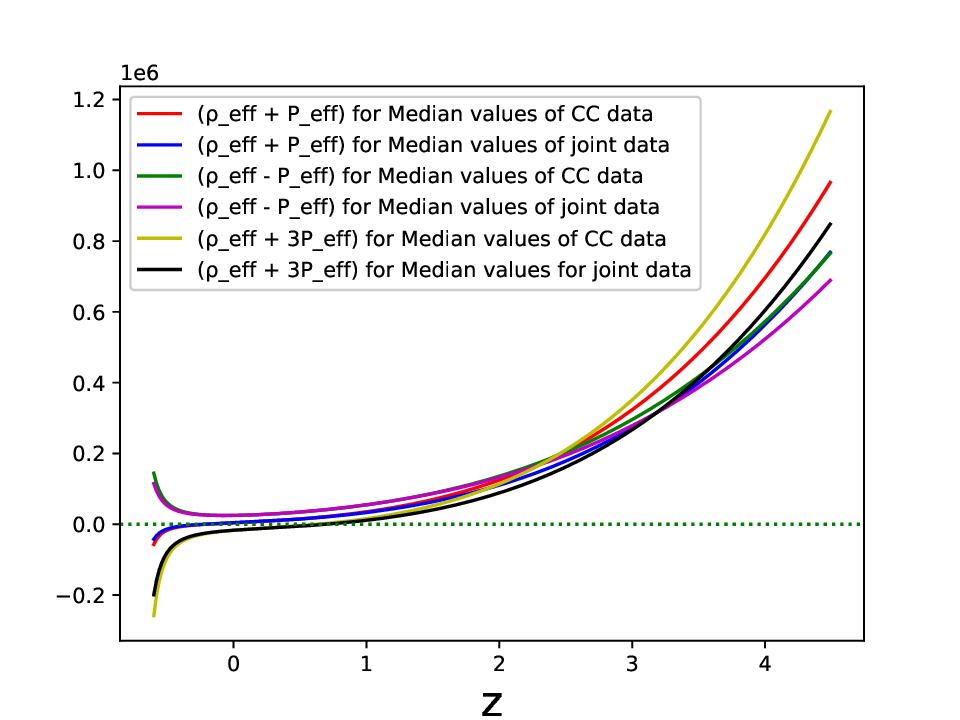}
\caption{ {\bf{For Model I:}} The components of energy conditions with $\mathit{z}$  }
\label{fig:10}
\end{figure}

 All energy conditions for Model-I are graphically illustrated in figures (\ref{fig:10}). From the graphical results in Fig. (\ref{fig:10}), it is evident that the Model-I satisfies NEC, WEC, and DEC upto present time. Because the reconstructed model shows current accelerated expansion, so the SEC (specifically $\rho_{eff}+3p_{eff}>0 $ is violated. As the Universe enters into phantom era during future, the NEC is violated, which may also lead to the violation of WEC and DEC. However, for the constrained values according to both data sets, the violation of $(\rho_{eff}+p_{eff}) \geq 0$ points for the presence of phantom kind of dark energy in the model. We may conclude that the phantom kind of dark energy may not be ruled out in this model. In the phantom dark energy dominated era, the violation of $\rho_{eff} +3p_{eff} > 0 $ will also be valid.
\par
\begin{figure}
	\includegraphics[width=17.5cm,height=8cm]{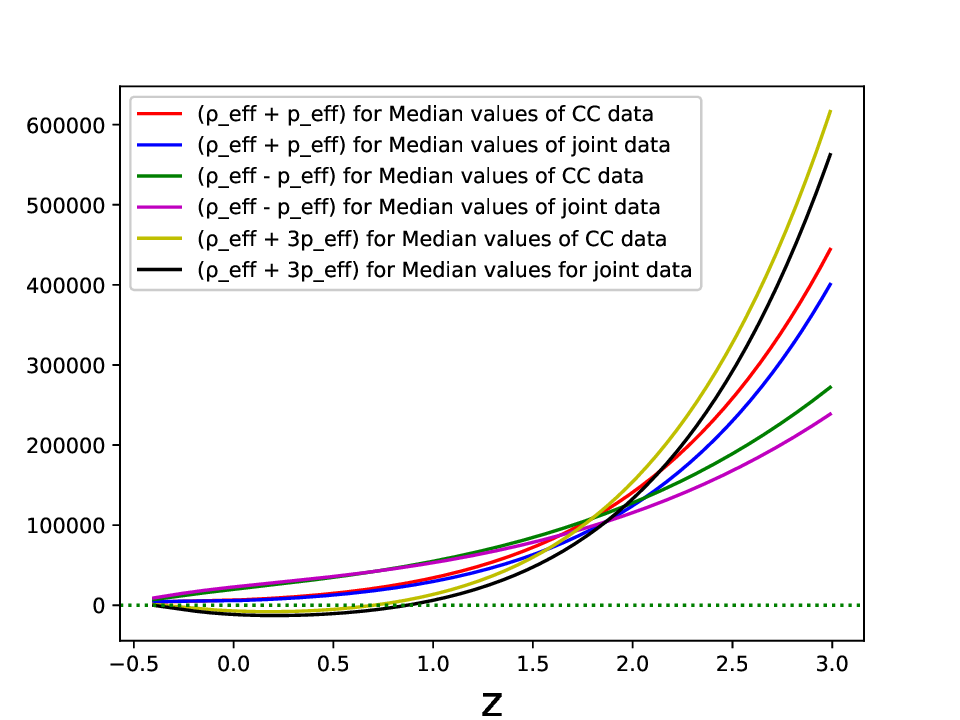}
	\caption{{\bf{For Model-II:}} The components of energy conditions with $\mathit{z}$  }
	\label{fig:10B}
\end{figure}

 For Model-II, graphical representations of all energy conditions are illustrated  in figure (\ref{fig:10B}). It is observed that $(\rho_{eff}+p_{eff})>0$, and $(\rho_{eff}-p_{eff})>0$ for the median values of model parameters. Through this observation, one can estimate that NEC, WEC, and DEC are satisfied for the model II as the universe evolves. Further, the trajectory of $(\rho_{eff}+3p_{eff})$ transits from positive to negative values.  Moreover, the SEC is satisfied at the early era but violates at present ($z=0$) in model II.

\subsection{Statefinder diagnostic}\label{sec:6.2}
\begin{figure}[!htb]
	\captionsetup{skip=0.4\baselineskip,size=footnotesize}
	\begin{minipage}{0.40\textwidth}
		\centering
		\includegraphics[width=8.0 cm,height=7.5cm]{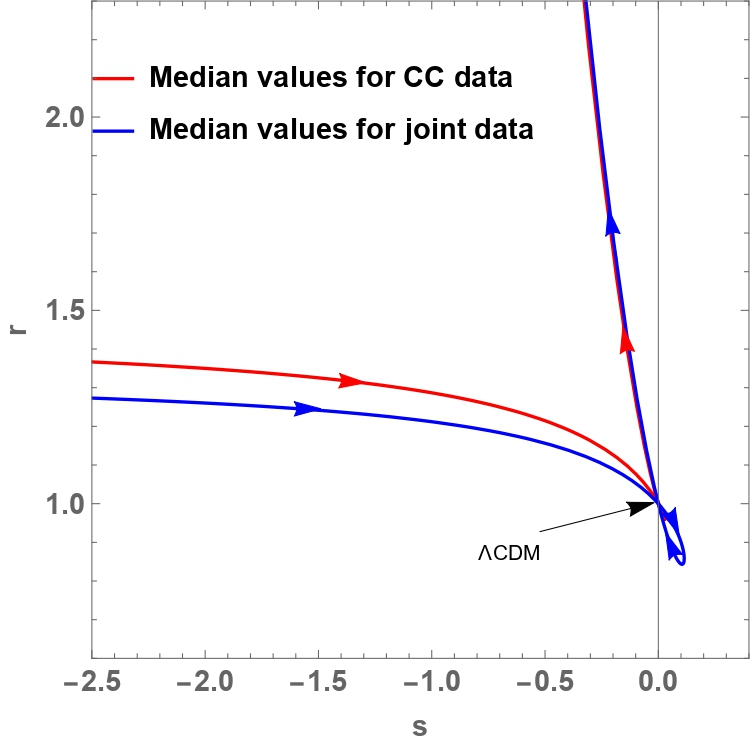}
		\caption{Model-I: Statefinder $(r, s)$ trajectory.}
		\label{fig:rs_model1}
	\end{minipage}\hfill
	\begin{minipage}{0.40\textwidth}
		\centering
		\includegraphics[width=8.0 cm,height=7.5cm]{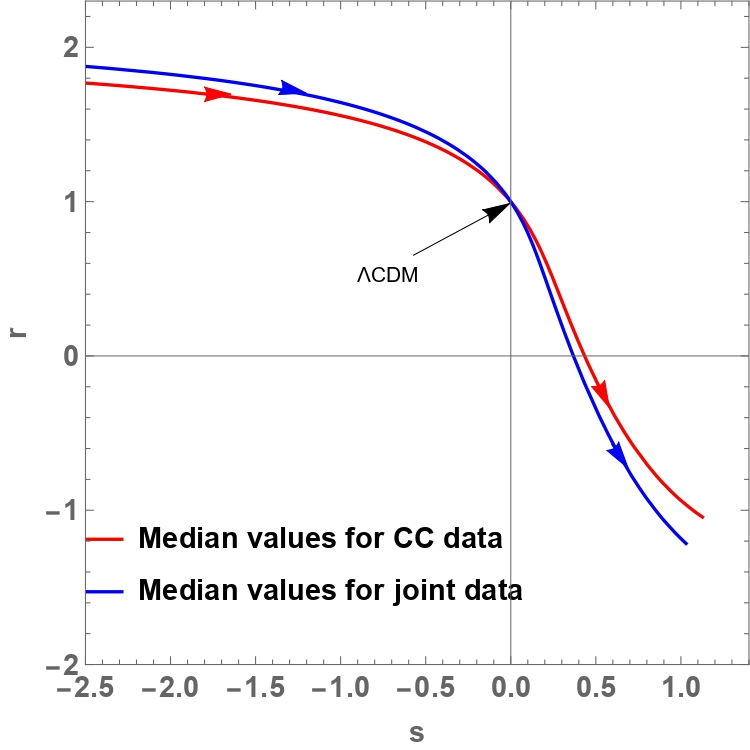}
		\caption{Model-II: Statefinder $(r, s)$ trajectory.}
		\label{fig:rs_model1B}
	\end{minipage}
\end{figure}

It is broadly acknowledged that the geometric parameters of the cosmological model may offer the profound intuitions for its dynamical behaviour. To explore different dark energy models other than to the standard $\Lambda$CDM  model, it becomes vital to analyse additional parameters other than the Hubble parameter and deceleration parameter. Higher-order derivatives of the scale factor $a(t)$ occurs as essential tools to define the dynamical characteristics of the universe. It leads us to discuss about the geometric parameters, which are consisting the higher derivatives of scale factor $(a)$. Sahni et al.~\cite{sahni2003statefinder, alam2003exploring} presented a pair of geometrical parameters, symbolized by $(r, s)$ and termed as the \textit{statefinder diagnostics}. The statefinder diagnostic is an effective tool for analyzing dark energy evolution in a model. The expressions for this diagnostic pair is given by:
\begin{align}
    r &= \frac{\dddot{a}}{a H^3}, \qquad s = \frac{r - 1}{3\left(q - \frac{1}{2}\right)}, \quad \text{where } q \neq  \frac{ 1}{2}.
\end{align}
Various dark energy (DE) models discussed in the literature can be characterized by different values of the statefinder pair $(r,s)$:
\begin{itemize}
\item For Chaplygin gas (CG) model, one may have $ (r>1, s<0)$.
\item The $\Lambda$CDM model correspond to $r=1, s=0$.
\item For Quintessence model, one may have $(r<1, s>0)$.
\item For Holographic dark energy model (HDE), $(r=1, s=\frac{2}{3})$.
\item For Standard cold dark matter (SCDM), one may have $ (r=1, s=1)$.
\end{itemize}
The figures (\ref{fig:rs_model1}) and (\ref{fig:rs_model1B}) illustrate the $(s, r)$  curve for Model I and Model II respectively. In figure (\ref{fig:rs_model1}), we see that for both data sets, the trajectory begins in the Chaplygin gas area during initial times, traverses the $\Lambda$CDM  point and ultimately develops by unifying the dark matter and dark energy in the model.\\
Figure (\ref{fig:rs_model1B}) illustrate that the evolution of the $(s, r)$ parameters initially start Chaplygin gas models, passes through the $\Lambda$CDM point and finally evolve like quintessence models in the present epoch for both data estimates.
\subsection{Age of the Universe}\label{sec:6.4}
The cosmic age $t(z)$ of a cosmological model can be calculated in terms of the red-shift $ \mathit{z}$ as\cite{tong2009cosmic}:
\begin{equation}
t(z) = \int_{z}^{\infty} \frac{dz}{(1+z)H(z)}.
\end{equation}
By using the Hubble parameter $ H(z) $ from equations (\ref{21}) and (\ref{103}), we calculate the present age of the universe for these models.\\
The age of the universe for model I is  $t_{0}$ (at $z=0$) = $13.22^{+0.165}_{-0.340}$ Gyr (for the CC dataset) and $t_{0} =13.41^{+0.14}_{-0.46}$ Gyr (for the joint data set). In  model II, the present age of the universe is $t_{0} = 12.74^{+0.18}_{-0.14}$ Gyr  and $12.62^{+0.33}_{-0.22}$ Gyr for CC and  joint data estimations respectively. We can see that the model's age of the universe has been change due to the absence of a structure formation era.
\section{Conclusions}\label{sec:7}
We study the universe evolution with the flat FLRW metric in $f(R,L_m)$ gravity where the equation of state parameter for dark energy evolves reciprocal to the redshift in one case and varies exponentially in another case. These EoS parameters are popularly referred to as the Gong–Zhang parameterization and are given by $\omega (z) =\frac{w_o}{z+1}$ and $\omega(z)=\frac{w_o}{z+1} \exp\left( \frac{z}{z+1} \right)$. The median value of model parameters are obtained using the MCMC analysis based on the observational data of CC and the supernovae type Ia Pantheon sample. The parameters are constrained through the MCMC analysis for both models and the results are summarized in Tables (\ref{table:1}) and (\ref{table:2}).
\par Moreover, the behaviour of deceleration parameters have been discussed. The evolution curve of the deceleration parameter reveals that the universe has transitioned from a decelerated to an accelerated expansion phase. For Model-I, the current values of deceleration parameters are $q_{0}=-0.60$ (CC) and $q_{0}=-0.59$ (joint) and the value of transition red-shift is $ z_{t} = 0.635 $ (CC) and $ z_{t} = 0.71  $ (joint). For Model II, the current value of deceleration parameter is obtain $q=-0.2718$ and $-0.4067$ for CC and joint data respectively. The value of transition red-shift is $ z_{t} = 0.7 $ (CC) and $ z_{t} = 0.871  $ (joint). The current value of the deceleration parameter is negative and it confirms the accelerated cosmic expansion in both models. These models are consistent with current observations of acceleration, but the early and late time dynamics may differ in these models. This is one of the important aspect of this study. The cosmological evolution in Model I progresses from early deceleration to the late-time super-exponential expansion. Unlike Model I, Model II exhibits no period of cold dark matter domination, resulting in the absence of a structure formation era. 
\vspace{0.2cm}\\
The equation of state $ \omega (z)=\frac{w_{0}}{1+z}\exp\left( \frac{z}{z+1}\right)$ tends toward zero in the early universe but does not exhibit a sufficiently prolonged CDM-like phase where $\omega \approx 0$ (corresponding to pressure less matter).  Model II predominantly exhibits accelerated behaviour at present. The limitation that structure formation processes such as the growth of cosmic web, galaxy clustering and formation of large-scale structures may not proceed efficiently in this model, as they typically require a prolonged matter-dominated phase where gravitational instability can amplify small initial density perturbations. However, the duration of matter-dominated phase are not sufficient. The model does not yield a sustained period where $\omega \approx 0$ and remains constant  which is essential for structure formation. Therefore, Model II lacks a prolonged, well-defined CDM-dominated era and this directly limits its ability to describe early structure growth. Without a CDM-like epoch, the gravitational collapse of matter overdensities is inefficient. This suppresses the formation of galaxies, galaxy clusters and the cosmic web. It leads to a mismatch with observed matter power spectra, halo abundances and large-scale structure surveys. The lack of CDM domination may suppress power on scales that entered the horizon during missing era, potentially conflicting with galaxy survey data.\\
This model II predicts the decelerating phase in the future, potentially resulting from energy exchange between dark energy and dark matter. Such a variation marks a key difference between both of these EoS parameterizations.\\
Additionally, we analyze the behavior of energy density and pressure. In Model-I, the energy density remains positive while the pressure starts with the positive values at high red-shift during the early universe and in current (and later epoch), it becomes negative. These findings are consistent with the accelerating universe’s expanding behavior.\\ 
The Figs. $(\ref{fig:4B})$ and $(\ref{fig:5B})$ illustrate the behavior of EoS parameter for Model I and Model II, respectively. In model I, the value of EoS parameter at present is  $ \omega=-0.801 $ and $\omega=-0.779 $ for CC and joint data estimates respectively. Similarly, for model II, $\omega=-0.768$ and $\omega=-0.822$ for CC and joint data. Hence, it is clear that from Fig. $(\ref{fig:4B})$ and $(\ref{fig:5B}$) that these models exhibits quintessence behavior for dark energy at present times.\\
For Model I, the obtained age of the universe is $t_{0} =13.22^{+0.165}_{-0.340}$ Gyr (CC) and $t_{0} =13.41^{+0.14}_{-0.46}$ Gyr (joint) and  for Model II, it is  $t_{0} =12.74^{+0.18}_{-0.14}$ Gyr (CC) and $t_{0} =12.62^{+0.33}_{-0.22}$ Gyr (joint). \\
The difference between behaviors of Model I and Model II may be due to the $\exp\left( \frac{z}{1+z} \right)$ term incorporated in Model II’s EoS parameter. Because of this term, the cold dark matter dominated era is absent in the model II. As a consequence, key cosmological parameters such as the deceleration parameter and cosmic age of model II differs from those in the standard $\Lambda$CDM model and the model I.
\section*{\textbf{Acknowledgements}}
Authors are thankful to the learned reviewer for the positive constrictive valuable suggestion. G. P. Singh and F. Rahaman are thankful to the Inter-University Centre for Astronomy and Astrophysics (IUCAA), Pune, India for support under Visiting Associateship program. R. Garg thanks to Dr. A. Singh for the insightful discussions. 
\section*{\textbf{ Data Availability Statement}}
All the publicly available data sources have been cited in the manuscript.
\section*{\textbf{ Declaration of competing interest}}
The authors declare that they have no known competing financial interests or personal relationships that could have appeared to influence the work reported in this paper.


\end{document}